\documentclass[sigplan,screen]{acmart}
\AtBeginDocument{%
  \providecommand\BibTeX{{%
    \normalfont B\kern-0.5em{\scshape i\kern-0.25em b}\kern-0.8em\TeX}}}

\acmConference[Conference acronym 'XX]{Make sure to enter the correct
  conference title from your rights confirmation emai}{June 03--05,
  2018}{Woodstock, NY}
\setcopyright{acmlicensed}
\acmPrice{15.00}
\acmDOI{10.1145/3622758.3622882}
\acmYear{2023}
\copyrightyear{2023}
\acmSubmissionID{onward23essays-p2-p}
\acmISBN{979-8-4007-0388-1/23/10}
\acmConference[Onward! '23]{Proceedings of the 2023 ACM SIGPLAN International Symposium on New Ideas, New Paradigms, and Reflections on Programming and Software}{October 25--27, 2023}{Cascais, Portugal}
\acmBooktitle{Proceedings of the 2023 ACM SIGPLAN International Symposium on New Ideas, New Paradigms, and Reflections on Programming and Software (Onward! '23), October 25--27, 2023, Cascais, Portugal}
\received{2023-04-28}
\received[accepted]{2023-08-11}

\usepackage{todonotes}
\usepackage{xcolor}
\usepackage{listings}
\usepackage{microtype}

\definecolor{codegreen}{rgb}{0,0.6,0}
\definecolor{codegray}{rgb}{0.5,0.5,0.5}
\definecolor{codepurple}{rgb}{0.58,0,0.82}
\definecolor{backcolour}{rgb}{0.95,0.95,0.92}

\lstdefinestyle{mystyle}{
    backgroundcolor=\color{backcolour},   
    commentstyle=\color{codegreen},
    keywordstyle=\color{magenta},
    numberstyle=\tiny\color{codegray},
    stringstyle=\color{codepurple},
    basicstyle=\ttfamily\footnotesize,
    breakatwhitespace=false,         
    breaklines=true,                 
    captionpos=b,                    
    keepspaces=true,                 
    numbers=left,                    
    numbersep=5pt,                  
    showspaces=false,                
    showstringspaces=false,
    showtabs=false,                  
    tabsize=2
}

\begin{document}

\title[Will Code Remain Relevant for End-User Programming with Generative AI?]{Will Code Remain a Relevant User Interface for End-User Programming with Generative AI Models?}

\author{Advait Sarkar}
\email{advait@microsoft.com}
\affiliation{%
  \institution{Microsoft Research and University of Cambridge}
  \city{Cambridge}
  \country{United Kingdom}
}
\affiliation{%
  \institution{University College London}
  \city{London}
  \country{United Kingdom}
}


\begin{abstract}


The research field of end-user programming has largely been concerned with helping non-experts learn to code sufficiently well in order to achieve their tasks. Generative AI stands to obviate this entirely by allowing users to generate code from naturalistic language prompts. In this essay, we explore the extent to which ``traditional'' programming languages remain relevant for non-expert end-user programmers in a world with generative AI. We posit the ``generative shift hypothesis'': that generative AI will create qualitative and quantitative expansions in the traditional scope of end-user programming. We outline some reasons that traditional programming languages may still be relevant and useful for end-user programmers. We speculate whether each of these reasons might be fundamental and enduring, or whether they may disappear with further improvements and innovations in generative AI. Finally, we articulate a set of implications for end-user programming research, including the possibility of needing to revisit many well-established core concepts, such as Ko's learning barriers and Blackwell's attention investment model.

\end{abstract}





\maketitle

\section{The Status Quo for End-user Programming Research}
\label{sec:eup-status-quo}

End-user programming (EUP) is the activity of writing a program for one's own use; one is both the programmer and the end-user of the program \cite{ko2011state}. This can be contrasted with software development, in which the programmers are typically not going to be people using the system. There is a huge variety of end-user programmers (EUPs) and end-user programming activities, such as writing spreadsheet formulas to analyse data, or a script to automate a daily workflow, or even programming as a hobby for personal creative satisfaction \cite{aghaee2015personality}.

Unlike professional software developers, EUPs typically have very little or no formal training in programming or computing. This gap in expertise is the primary challenge faced by EUPs trying to achieve their task (though there are other differences between EUPs and professionals, such as their motivations for programming, strategies for learning, debugging, etc. which add further complexities).

End-user programming research thus aims to bridge this gap. So far, its aim can be described as improving the ability of EUPs to use \emph{formal systems}. The term ``formal systems'' is a shorthand to mean a predictable and deterministic interface which may include a programming language (e.g., spreadsheet formulas, or block programming in Scratch \cite{resnick2009scratch}) and other interface elements (e.g., the spreadsheet grid).

EUP research helps users with formal systems in three ways. The first approach is to help users \emph{learn} how to use formal systems. This concern is shared by computer science education research (CSER) \cite{fincher2004computer}, and includes research on novice programming, the design of programming tutorial aids, multiple representations systems \cite{stead2014learning}, etc. \nocite{sarkar2018people}

The second approach is to reduce or scaffold the expertise required to use formal systems. This is the motivation behind many visual programming languages, such as Scratch~\cite{resnick2009scratch}, which uses blocks and slots of various shapes to anticipate and prevent syntactic and type errors, since a block of a certain shape can only fit into a corresponding slot.

The third approach aims to reduce the reliance on coding, such as programming by example (PBE) \cite{lieberman2001your} and programming by demonstration (PBD) \cite{cypher1993watch}. Thus, by showing examples of input and output, or manually demonstrating a particular process, users can define intended behaviour (code generation is often facilitated by a method such as program synthesis \cite{manna1971toward}). While the aim is to reduce the reliance on coding, in practice many PBE/PBD systems do not entirely avoid interaction with the generated code. For instance, Flash Fill \cite{gulwani2011automating}, a commercially implemented PBE system for string manipulation formulas in spreadsheets, was initially deployed without showing the generated programs to the user, but was later extended to show the user its generated formulas (in some cases) for explanation, verification, and debugging.

In summary of the status quo: EUP research aims to help users avail of formal programming systems by improving their ability to learn, to reduce the expertise requirements of formal systems, or by offering alternatives to writing code.

\section{The Scope of the Term ``Generative AI'' for End-User Programming}

The term ``generative AI'' is extremely broad, encompassing many types of system with different capabilities, sometimes referring to a core algorithm (such as the transformer \cite{vaswani2017attention}), a particular instantiated model (such as GPT-4 \cite{openai2023gpt4}; these are also sometimes referred to as ``foundation models'' \cite{bommasani2021opportunities}), or a productised system which may be comprised of an ensemble of multiple models together with prompt engineering, safety heuristics, and user interface affordances (such as ChatGPT\footnote{\url{https://openai.com/blog/chatgpt} (last accessed 23 June 2023)}). In this respect ``generative AI'' has taken on a similarly pluralistic nature as terms such as ``machine learning'' and ``artificial intelligence'' which can span an immense range of tools and techniques \cite{mackenzie2017machine}. 

The additional descriptor ``generative'' is a reference to the fact that these models can \emph{generate} information artefacts (such as images or text), by modelling a high-dimensional space from its input data from which new points can be sampled. This is as opposed to discriminative models, which can classify, label, score, or transform their input based on training examples, but do not explicitly model the input space \cite{lasserre2006principled}. It must be noted that while the term ``generative AI'' has only come into widespread usage in 2023 and is used to refer to contemporary generative models, research into generative models long predates this \cite{revow1996using, everitt1984introduction}, and ``latent variable models'' were proposed as early as the turn of the 20\textsuperscript{th} century \cite{Cai2012-kk}.

To focus the scope of the discussion in this paper, it is worth defining ``generative AI'' in terms of the aspects which are of greatest interest to EUPs, as well as the aspects which differentiate ``generative AI'' from previous generations of AI tools which have also been applied in the EUP context.

First, we are talking about \emph{tools} which are directly used by EUPs. An EUP does not interact with an algorithm, or a specific model, but a tool which may consist of multiple models that are part of a broader system of heuristics, prompt engineering, and user interface elements. From the perspective of the end-user experience, there are many aspects of these heuristics and interface elements which impact their ability to use the underlying model.\footnote{There are also elements that impact usability but have little or no connection with the underlying model, such as the text in dialogue boxes and the placement of menu buttons, which are not of concern here.}

Second, we are talking about tools which are used in a \emph{programming} context. There are many ways to define programming, but a particularly useful and influential definition in end-user programming research is that programming is any activity exhibiting ``the property that the user is not directly manipulating observable things, but specifying behaviour to occur at some future time'' \cite{blackwell2002programming}. Generative AI tools can be applied in a wide variety of contexts, and many of them relate to the direct production of artefacts (text, images, etc.) which are not interpreted as specifying behaviour to occur in the future -- we exclude these from consideration.

Finally, we are talking about tools built on \emph{contemporary} machine learning techniques as of this writing in 2023. Despite the fact that generative models have a long history, generative AI models, such as large language models \cite{brown2020language} and image generation models \cite{ramesh2021zero} have recently seen a significant step-change in capabilities. The reason for this is a combination of advances in hardware (such as GPU clusters for training), algorithms (such as the transformer architecture), and the availability of Internet-scale datasets \cite{sarkar2022programmingai}. This has enabled models to achieve human-level performance for the first time in a wide variety of benchmarks including code generation, speech recognition, image generation, even passing the bar exam \cite{katz2023gpt}. This is the latest development in a period typically dated to begin in 2016 that has been described as the ``third summer'' of AI,\footnote{2016 is the year that Deepmind's AlphaGo program \cite{silver2016mastering} beat grandmaster Lee Sedol, though some scholars point as early as 2012, referring to the so-called ``ImageNet moment'' referring to the solution of the ImageNet challenge by deep convolutional networks \cite{krizhevsky2017imagenet}, while others look as recently as 2018, referring to the BERT model which similarly swept natural language processing benchmarks and is referred to as ``NLP's ImageNet moment'' \cite{devlin2018bert}. A detailed historiography of the periodisation of AI research is out of scope.}  following a common periodisation of AI research as measured by ``rapid scientific advances, broad commercialisation, and exuberance'' \cite{kautz2022third}. A relatively stable term of art accepted and advocated within the AI research community that encapsulates the advances of the third summer is ``deep learning'' \cite{lecun2015deep}, which is broad enough to encompass a variety of approaches developed in recent years while being specific enough to exclude older generative approaches.

Putting these considerations together, the following definition is adopted for the remainder of the paper: 
\begin{quote}
    By generative AI, we mean \emph{an end-user tool, applied to programming, whose technical implementation includes a generative model based on deep learning}.
\end{quote}

Thus, some concrete examples which fit the definition and scope of generative AI in this paper are: GitHub Copilot\footnote{\url{https://github.com/features/copilot} (last accessed 25 April 2023)}, an end-user tool which offers code completion assistance based on the OpenAI Codex large language model; end-user tools for naturalistic language-based programming in spreadsheets such as those built by Liu et al. \cite{liu2023what} (also using Codex); and ChatGPT when it is being used to generate code, debug code, or explore APIs and documentation.

Some concrete examples which do \textit{not} fit the definition and scope of generative AI in this paper are: CodeT5 \cite{wang2021codet5}, a code generating large language model (uses deep learning, is programming-oriented, but is not an end-user tool); early programming-by-demonstration systems such as Allen Cypher's Eager \cite{cypher1991eager} (is programming oriented, is an end-user tool, but does not use deep learning); and ChatGPT when it is being used to generate a short fiction story (uses deep learning, is an end-user tool, but is not supporting a programming activity).

\section{Motivations for Applying Generative AI to End-User Programming}
\label{sec:motivations}

Why are the capabilities of generative AI important or beneficial for EUP, and how does it differ from existing approaches? As we have seen, EUP research is concerned with helping users achieve programming tasks by improving learning, reducing expertise requirements, and reducing reliance on coding. AI approaches prior to deep learning have been applied in various ways to each of these aims, such as generating and recommending tutorials \cite{zhang2010towards,klavsnja2011integration}, detecting errors and suggesting fixes \cite{hermans2015detecting}, and the generation of code from demonstrations \cite{cypher1993watch}.

In several EUP activities, generative AI creates improvements in \emph{degree} over previous approaches. That is, it improves performance in terms of scope and accuracy. For example, let us consider five categories of EUP activity in which generative AI creates an improvement in degree: authoring, debugging, reuse, comprehension, and learning.

\textbf{Authoring.} The principal activity of EUPs is authoring, that is, actually writing code. Quantitative evidence for the improvement in degree afforded by generative AI is given by tracing the performance of various approaches in coding benchmark tests \cite{austin2021program, touvron2023llama, jiang2023selfevolve}, where it is now possible to solve a much wider variety of programming tasks using natural language descriptions of the problem, and success rates for programming tasks described as being representative of ``entry-level programmer'' skill now regularly approaches 80\%.

\textbf{Debugging.} Another key activity of EUPs is in debugging: detecting and fixing errors in existing code. Here again, quantitative studies and benchmarks of error detection and fixing show a significant improvement in degree in comparison to previous approaches \cite{kang2023explainable, fan2022improving}.

\textbf{Reuse.} Often, EUPs do not seek to directly author code starting from a blank slate, rather, they seek to reuse code available on the Web, code they have previously written, or code written by colleagues or stored in institutional repositories \cite{srinivasa2016foraging, lau2021tweakit}. There are several key challenges in the reuse activity, principally, locating relevant code, evaluating its suitability for reuse, and making necessary modifications to make the code fit their current context. Studies have found that even with interactive support for code reuse, EUPs can struggle to adapt code due to a lack of formal programming expertise \cite{lau2021tweakit}. Generative AI tools excel at facilitating reuse and adaptation of code, and mark a significant improvement in degree in comparison to previous approaches \cite{sarkar2022programmingai}.

\textbf{Comprehension.} EUPs often seek to understand code for reasons other than debugging. For example, users receiving an unfamiliar spreadsheet from a colleague may review formulas in the spreadsheet to understand the underlying data sources for a particular value in their spreadsheet which is relevant to a decision they need to make \cite{srinivasa2021spreadsheet,sarkar2022end}. Studies have shown that generative AI has significantly improved capabilities in code explanation and summarisation for students and non-experts (though challenges remain) \cite{kang2023explainable, macneil2023experiences}.

\textbf{Learning.} A wide variety of AI approaches to facilitate learning for EUPs have been previously explored, from suggesting relevant tutorials \cite{zhang2010towards,klavsnja2011integration}, to proactive suggestions for improving the likelihood of EUPs to write test cases to improve the robustness of their programs \cite{wilson2003harnessing}, and overcoming conceptual learning barriers \cite{jernigan2015principled}. However, a key challenge remains in adapting tutorials and learning materials to an individual's problem context (what exactly are they trying to solve), their existing knowledge, and their learning style \cite{burnett2016gendermag, sarkar2023should}. Studies of spreadsheet learning show that learning from colleagues or ``hallway experts'' is common because it is an ideal learning situation: the colleague understands the user's problem context and can tailor an explanation precisely to the learner's needs \cite{sarkar2018spreadsheetlearning}. This ``holy grail'' of personalised learning is another area where generative AI has shown to provide a significant improvement in degree, where explanations can be generated on an ad hoc basis for arbitrary code with arbitrary amounts of detail \cite{macneil2022automatically, denny2023computing}, and it is likely that such explanations can be tailored with much greater precision to a particular learner in a particular instance.


Besides improvements in degree, generative AI also creates improvements in \emph{kind}, that is to say, it can broadly shift EUP activities in novel directions, or enable kinds of assistance that were not possible with previous approaches. For example, let us consider the cases of one-off automations and exploratory programming.

\textbf{One-off automations.} Recall our working definition of programming as ``specifying behaviour to occur at some future time''. This may suggest that EUPs write programs much as software developers do, as reusable assets which can be used repeatedly over time on different input data. Certainly many end-user automations are like this (consider the formulas in organisational spreadsheets which are often highly ``templatized'', reused broadly and longitudinally across the organisation, sometimes for many years \cite{hermans2011supporting}). However, a lot of EUP is \emph{not} like this, where instead the objective is to develop ad-hoc data processing or cleaning scripts which are used exactly once for a specific transformation and then discarded. In such one-off automations a lot of interesting end-user programming behaviour is observed, such as accepting errors or brittleness in the program, and transforming the data through a patchwork of manual and automated steps \cite{pandita2018no,sarkar2023should}. Generative AI technologies are capable of transforming, cleaning, and augmenting data \emph{directly}, which would eliminate the need for writing a reusable script for one-off automations entirely \cite{vos2022towards, jaimovitch2022can}. This would shift the scope of EUP activities away from writing such automations, but it would also create new challenges for EUPs, for instance in verifying that such transformations have been done correctly, and in making the occasional transition from a fundamentally opaque transformation performed by a generative AI model to a more repeatable and well-understood script.

\textbf{Exploratory programming.} Finally, many EUP activities involve an exploratory aspect; where the programming objective is not known \emph{a priori} but discovered through trial and error and experimentation \cite{kery2017exploring}. This is not just true of artistic applications of programming such as the generation of digital art or live coding music \cite{collins2003live}, where the exploration may not converge to a formally ``correct'' solution (rather the exploration converges to a state of subjectively assessed completion), but is also true of situations such as exploratory data analysis where the rough forms of acceptable solutions may be known beforehand, but the precise procedures are still to be determined \cite{morgenthaler2009exploratory}. Here, generative AI not only increases the rapidity of such exploration but also enables forms of assistance that were not possible with previous approaches, such as the generation of alternative narratives to trigger nonlinear ``leaps'' in the user's thinking (though this is yet to be investigated in an EUP setting) \cite{singh2022hide, sarkar2023exploring}. \nocite{chalhoub2022s, joharizadeh2020gridlets}


\section{The Potential Intensification of End-User Programming: The Generative Shift Hypothesis}

Generative AI can be applied to generate code in a traditional programming language based on natural (or naturalistic) language prompts. A brief overview of large language models for code generation is given in Sarkar et al. \cite{sarkar2022programmingai}. This technology has already been commercialised in a number of code editors and extensions, such as GitHub Copilot.\footnote{\url{https://github.com/features/copilot} (last accessed 25 April 2023)}

As a method for enabling people to program without writing code directly, generative AI can be viewed as an evolution of previous methods such as PBE, PBD, and older syntax-directed or machine-learning based code autocompletion.

In other ways, as explained in Section~\ref{sec:motivations}, the maturation of generative AI may herald a revolution in capabilities, in particular for EUPs. This is what we will refer to as the \emph{generative shift hypothesis}: a radical widening in scope and capability of EUP due in particular to increasing use of generative models. 

The generative shift hypothesis posits the following qualitative and quantitative shifts:

\begin{itemize}
    \item EUP will be applied \emph{more intensively} to existing tasks, with more sophisticated and deeper automation being applied to scenarios which are already sites of EUP (such as spreadsheets).
    \item EUP will be applied in \emph{more contexts} and to more tasks than before, which were previously not sites for the widespread application of EUP. This includes scripting of tasks across applications and across data sources (this is related to the enterprise concept of ``robotic process automation'').
    \item EUP will be applied \emph{more frequently}. The vastly reduced costs of generating code from naturalistic utterances will shift the attention investment \cite{blackwell2002first} balance, making it more practical to attempt to automate tasks more often.
\end{itemize}

It is not within the scope of this essay to gather evidence for or against the generative shift hypothesis, and assess whether, and how fast, it might happen. Rather, we take the position of assuming it will happen, and attempting to analyse the role of formal systems in such a future.

Effectively guiding a generative model to produce the desired outcome is an area of active research (e.g., \cite{liu2022design,white2023prompt,strobelt2022interactive}). Current commercial tools can generate code from naturalistic language ``prompts'', or auto-complete partially written code, or a combination of both. For non-programmers, the main mode of interaction is via language. This opens the possibility for EUP to become an activity that does not rely on learning or engaging with a formal programming system at all.

Generative AI therefore raises a critical, perhaps existential question for EUP research: what is the role of the current research agenda which aims to help users learn formal systems, or reduce the expertise requirements of formal systems? In a world where generative models are highly performant at translating naturalistic language into code, is there still a role for learning and interaction with formal systems for EUPs? Put simply: \emph{will code remain a relevant user interface for end-user programming}?


\section{Does Code Still Matter? Evaluating the Value Propositions of Formal Systems}

To answer whether code may become irrelevant for EUPs, it may help to enumerate some current reasons that using a formal system, or combining a formal system with generative AI, might still result in a superior user experience than interacting purely through informal language. This section will discuss the following, possibly incomplete, list of reasons why code still matters: limited abilities, control, agency, awareness of the possible, explanation and interpretation, debugging, trust, and the illusion of informality. We will also consider the extent to which each value proposition might endure or become less relevant, as generative AI improves.

\textbf{Limited abilities.} The first reason is simply that generative AI can fail to correctly ``translate'' a natural language prompt into a suitable and correct program. It is difficult to draw clear boundaries around what these models can and cannot do. For one, they are opaque and non-deterministic, and can produce widely divergent responses to the same query. For another, the response is heavily influenced by the prompt, the training data, hyperparameters, and any other heuristics or processing applied to the input or output in a particular system. Finally, the state of the art is rapidly evolving. Thus, it would not be appropriate to make general statements such as \emph{``generative AI cannot currently solve problem X''} because it makes a universal claim about the capabilities of every system, with their own ensemble of models, heuristics, training data, etc. Rather, every system's response is fundamentally unique and varies depending upon the prompt, heuristics, and hyperparameters.

With these caveats in mind, it is worth looking at a small set of examples of generative AI failure, only to illustrate the kinds of problems that are not \emph{typically} solved by end-user tools as of this writing in 2023.

\begin{lstlisting}[float, language=Python, label=lst:1, caption=Python code generation failure example from Denny et al. \cite{denny2023conversing}.]
# Given a list of strings and a string s, 
# return the average length of all strings 
# containing s.

def averageContainingStringS(strings, s):
    count = 0
    for string in strings:
        if s in string:
            count += 1
    return len(s) * count / len(strings)
\end{lstlisting}

\begin{lstlisting}[float, label=lst:2, caption=Code generation failure example from Khatry et al. \cite{khatry2023words}.]
Prompt: "Select all rows where the entry in column 
'gamma' is less than 40 and select all rows where 
the entry in column 'gamma' is more than 53"

Response:
Table.SelectRows(#"Table1",each [gamma]<40)
\end{lstlisting}

\begin{lstlisting}[float, language=Python, label=lst:3, caption=Python code generation failure example from Liu et al. \cite{liu2023what}.]
Prompt: "How many super bowls has New Orleans won"

Response:
df[df['Host City'] == 'New Orleans']['Winner'].count()
\end{lstlisting}

The first example (Listing~\ref{lst:1}) comes from a paper exploring the limitations of GitHub copilot for CS1 problems \cite{denny2023conversing}, where copilot generates an incorrect function for calculating the average length of strings in a list, by generating nonsensical arithmetic. The second example (Listing~\ref{lst:2}) comes from Khatry et al. \cite{khatry2023words} and shows another simple example where the generated code satisfies one of the requested conditionals but ignores the other.

The final example (Listing~\ref{lst:3}) is drawn from a paper applying code generation models to data analysis problems in spreadsheets, where the full problem context can be seen \cite{liu2023what}. In this case, the code has incorrectly selected the ``Host City'' column to compute a count of relevant strings, whereas it should have selected the ``Winner'' column. Liu et al. \cite{liu2023what} characterise several types of failure modes with examples, such as the generation of non-executable code, selection of incorrect input columns, incorrect output formats (e.g., the generated code overwrites a column instead of generating a new one), or generating raw data output rather than a calculation.


As these examples suggest, there are situations where generating the required code is either beyond the capabilities of generative AI, or for some reason or another incorrect code is generated even if it is within current capabilities. To cope with such situations, EUPs might need to engage with a formal system.

On the other hand, at the time of writing in early 2023, we are in a transitional moment with generative AI technology. It is clear that generative models will continue improving due to the current strategy of increasing the scaling of parameters and training data.\footnote{This trend is almost certainly not going to continue indefinitely; we are approaching limits of computation and data availability, and others have pointed out various limitations to the so-called ``scaling hypothesis'' \cite{villalobos2022will,sevilla2022compute,kaplan2020scaling,sarkar2022explainable}, but a discussion of this is out of scope.}

Therefore, the problem of limited abilities, by definition, may seem to ameliorate as generative models improve. However, as the capabilities for automation improve, so might the demand for automation; it would be a fallacy to assume that EUP demands are stable over time and that today's EUP tasks are representative of future ones. Transport planners often engage in road-widening schemes to alleviate traffic congestion, only to find that the increased capacity is immediately consumed by increased demand \cite{metz2021economic}. In the 19\textsuperscript{th} century, economist William Jevons observed that improvements in the efficiency of coal engines paradoxically resulted in an increased demand for coal \cite{alcott2005jevons}. These are examples of ``rebound effects'', where gains in efficiency, capacity, and capability, are offset by changes in human behaviour. Similarly, the improvement of generative AI in addressing basic tasks may result in end-user demand for even more sophisticated and nuanced automation, which again shifts the goalposts for AI. On the other hand, EUPs already find engaging with formal systems a challenge; it is therefore difficult to see how engagement with formal systems might be a suitable fallback in the nuanced cases where generative models of the future fail.


\textbf{Control.} The second reason is to give direct and nuanced control over the behaviour of the system. Composing a program using a formal system allows users to directly express both ``what'' to compute and ``how'' to compute it. It allows EUPs to directly implement desired operations, as well as avoid undesired ones (e.g., avoiding a very slow API call, or avoiding a lookup of private data if it is unnecessary).

Formal systems offer direct and nuanced control, but varying degrees of control can also be achieved using informal systems. For instance, faceted natural language prompts which specify input and output types, or decompose a problem into smaller steps, offer increased control while retaining the relative informality of natural language \cite{mishra2021reframing, liu2023what}. As generative models improve in performance, providing control without resorting to formal systems will be an important target for the EUP research agenda.

\textbf{Agency.} Related to control, the third reason is agency \cite{coyle2012did}, a term from cognitive neuroscience referring to the feeling of being an active agent able to effect change in the world. A sense of agency improves the EUP experience, a lack of agency worsens it. Learning to code in a formal system has been associated with a sense of agency, and the related concept of computer self-efficacy \cite{beckwith2006tinkering}.

It is important to note that while control and agency are related, they are distinct constructs. Control refers to the perception of influence over the external environment, whereas agency focuses on the perception of being an intentional actor who can initiate and execute actions. Both control and agency play crucial roles in human experience, motivation, and well-being.

Though they are often interdependent, it is possible to have agency without (some types of) control. For example, \emph{``agency is influenced by control specified at different hierarchical levels. [...] when higher-level control is exercised (i.e., goal-level control) lower level control processes (i.e., perceptuo-motor control) have no influence on sense of agency''} \cite{limerick2014agency}. Conversely, it is also possible to have (some types of) control without agency. For instance, studies of assisted mouse cursor pointing (an example of a lower-level perceptuo-motor process) have shown that \emph{``beyond a certain level of assistance users experienced a detectable loss in their sense of agency''}, even when the user had full goal-level control (i.e., the final destination of the cursor) \cite{coyle2012did}. 

An interesting question posed by the generative shift for interface design, therefore, is what types of control over the generation process are necessary or sufficient for creating a sense of agency. Direct control via a formal system can create a sense of agency, but just as there are alternative ways of offering control, so there are alternative ways of fostering agency. At one extreme, the sense of agency can be manipulated using placebo controls (such as the notorious pedestrian traffic light buttons that serve only to pacify impatient pedestrians and have no effect on the scheduled changes of the lights \cite{middaugh2018delusion}). Moreover, user agency is a complex, multi-level phenomenon: the overall agency of an EUP might increase with the increased automation capabilities offered by generative AI, even if they find it difficult to control the behaviour of the AI in particular instances due to the lack of a formal system. As with control, fostering and maintaining an appropriate sense of agency will be an important target for the EUP research agenda.


\textbf{Awareness of the possible.} The fourth reason that learning and using a formal system is still useful is the ``awareness of the possible'' \cite{sarkar2023should}; experience with a formal system gives users a thought language within which to define problems and envision potential solutions. Knowing \emph{that} something can be automated is as important as knowing \emph{how} to automate it; a user with a powerful generative AI that can automate some task is still at a disadvantage if they do not conceive of that task as something which can be automated. 

This benefit of formal systems may be surpassed by improved generative AI as well as other interaction design techniques. Generative models may be able to suggest potential suitable automations and different strategies for implementing them, educating the user along the way. For example, in the field of data analysis, the aim of Intelligent Discovery Assistants (IDAs) has been to support users to form a strategy for achieving some analysis goal \cite{serban2013survey}; similarly research into exploratory programming \cite{kery2017exploring} aims to support users in shaping their goals through experimentation. These fields have produced general interaction techniques for visualising and interacting with a possibility space which may be applied to interaction with generative models, without requiring end-users to engage with code. As with the previous points, facilitating this awareness with generative models will also be an important target for EUP research. \nocite{sarkar2016visual}

\textbf{Explanation, interpretation, and debugging.} Research has shown that explaining the behaviour of AI-driven EUP systems is desirable for many reasons \cite{kulesza2015principles}; allowing EUPs to engage with the formal system(s) in which a generative model might emit scripts is one way (albeit not the only way) to provide explanation and interpretability. \nocite{gorinova2016live}

Debugging is closely related to explanation and interpretation. Engagement with formal systems is a potentially superior way for EUPs to find and fix errors, than engaging with a generative model purely through natural language. For example, if the script generated by a generative model contains an off-by-one error, viewing the generated code might make it significantly easier to identify and correct it.

Formal systems offer a predictable route towards explanation, interpretation, and debugging. Yet each of these already presents a formidable challenge for EUPs, indeed much EUP research focuses on helping EUPs carry out each of these activities by scaffolding formal systems to overcome their limitations (e.g., the ``idea garden'' \cite{jernigan2015principled}). The design challenge is likely to increase further as generative models improve at synthesizing more sophisticated programs. We probably want different approaches to explanation which do not rely on engagement with formal systems; EUP and interactive machine learning research has already been developing various techniques that may be suitable starting points \cite{kulesza2015principles,ko2004designing}.

\textbf{Trust.} Formal systems allow EUPs to verify the behaviour of the generated code in terms of well-understood and well-trusted procedures. For instance, when a spreadsheet user sees that a sum is calculated with the spreadsheet \texttt{SUM} function, they trust that it is correctly computing the sum because it is a deterministic algorithm written and checked by a trusted party (e.g., Microsoft or Google). Inspecting the formal output of a generative model may therefore be an important route towards trusting the result. It is worth noting here, though, that code ``correctness'' is not as important to EUPs as it is to professional software developers; research has shown that EUPs commonly use incorrect or incomplete code if it helps them partially automate their workflow, and are happy to manually correct errors, or accept errors which they see as irrelevant to the task at hand \cite{pandita2018no}.

Expressing a computation in terms of trusted and verified operations (such as \texttt{SUM}) is an important route for establishing trust in the correctness of generated code. This strength of formal systems is unlikely to disappear simply by improving the performance of current models. Generative models can be asked to explain code in natural language, but they are prone to hallucination, as their output represents statistically likely completions rather than true and verifiable statements \cite{ouyang2022training}. In the future, it may be possible to constrain explanations of a script to refer to provably correct and verifiable procedures. To pursue this strategy and develop usable EUP systems along these lines will require a collaborative effort between researchers skilled in generative AI, programming languages, and human-centric end-user programming.


\textbf{Illusion of informality.} The eighth and final reason is slightly unlike the others, and is rather a meta-level observation about the nature of using generative models. The observation is that the disappearance of formality may be an illusion; generative models still require high levels of craft expertise to use effectively, and the shift to ``prompt engineering'' hasn't eliminated programming at all, but simply shifted it into a higher level of abstraction. This is a perspective harboured by some (though not all) professional software developers who use generative AI assistance in their work \cite{sarkar2022programmingai}.

As generative models get better, will ``prompt engineering'' remain a form of skilled craft practice that is distinct from ordinary communication skills, or will generative models reach a point where they can interpret truly arbitrary natural language at or exceeding human proficiency? On one hand, Miller and others have pointed out that a major hurdle for AI to interpret the meaning of a query is understanding the context of that query \cite{miller2019explanation, sarkar2022explainable}, with the implication that future models, which will become better at including or inferring this context, will allow for more informal querying styles. On the other hand, the imprecision of natural language, particularly when it comes to discourse on matters of logic, mathematics, philosophy, and science, has long been seen as a major drawback and spurred many attempts to design more logically ``perfect'' languages \cite{Eco1997-sa, okrent2009land}. Indeed, the program of analytic philosophy which was born out of such concerns eventually gave rise to our modern programming languages \cite{arawjo2020write}, and it is interesting that programming via generative models brings us back, full circle, to natural language. However, the trend of improving generative models seems to imply that rather than ``prompt engineering'' remaining just like programming but at a higher level of abstraction, the application of language formality for precision, brevity, etc. in future systems will become more social and discretionary. Language for instructing generative models may develop much as scientific or legal language develops as a way for scientists and lawyers to communicate more effectively: organically, through power relations enacted by individuals, groups, and institutions \cite{Foucault1966-ju}.

\section{Limitations and Challenges Posed by Generative AI in End-User Programming}
\label{sec:limitations}

There are many limitations and challenges of applying generative AI in end-user programming. Many of the challenges and ethical issues around trust, verification, privacy, bias, credit, and accountability are not unique to end-user programming: they apply to all applications of AI to software development, and at the broadest level they apply to all applications of AI. 

The focus of this paper is on the role of formal systems in the future of end-user programming, and a detailed discussion of ethical challenges would diverge from this focus. These have been extensively reviewed in recent research \cite{weidinger2021ethical, solaiman2019release, tamkin2021understanding}. Nonetheless, it is worth briefly acknowledging these challenges to highlight that the generative shift is not an uncontestedly positive improvement for end-user programmers.


\textbf{Problems for CS education.} There is concern that generative AI may undermine the learning of novice programmers \cite{denny2023computing, becker2023programming}. Besides issues of academic integrity and misconduct, educators are apprehensive that over-reliance on generative AI may hinder the development of novice programmers' metacognition and encourage what is considered ``bad habits'' in code.

\textbf{Errors.} Generative AI systems may not always produce high-quality outputs, and the generated outputs may contain errors or inconsistencies that are hard to detect or correct by end-user programmers \cite{sarkar2022programmingai}.

\textbf{Trust and transparency.} Generative AI systems may not be able to explain how they generate their outputs, or why they choose certain outputs over others \cite{sarkar2022explainable}. This may make it difficult for end-user programmers to trust or verify the generated code, or to understand its logic and functionality. They may not provide any documentation or comments for the generated code, or provide incorrect documentation, or may not allow the end-user programmer to modify or customise the code.


\textbf{Privacy and security.} Code generated by generative AI systems may collect, store, or use personal or sensitive data from end-users or other sources, such as health records, biometric data, or financial data. This may raise concerns about the protection of data privacy and security, and the potential for data breaches, misuse, or abuse. They may access or expose confidential or proprietary information from end-users or their clients \cite{harrer2023attention}.

\textbf{Misinformation.} Generative AI systems may generate outputs that are false, misleading, or deceptive, either intentionally or unintentionally. Generative AI models can produce outputs that are coherent and convincing, but not necessarily accurate or factual. They can also invent references and sources that do not exist, or contain biases from the training data. This can mislead or confuse end-user programmers who rely on generated code for their own use. For example, a generative AI tool may generate a code snippet that performs a data analysis, but the results may be incorrect or skewed by the underlying data or model assumptions. This may create risks of misinformation and manipulation, and undermine the trustworthiness and credibility of information sources. \nocite{tafur2023user}


\textbf{Bias and discrimination.} Generative AI systems may generate outputs that are unfair, biased, or discriminatory, either intentionally or unintentionally. They may generate code that reflects or amplifies existing social biases or stereotypes, such as gender, race, or ethnicity \cite{bender2021dangers}. This may result in harms or injustices to individuals or groups, such as exclusion, marginalisation, or oppression.

\textbf{Accountability.} Generative AI systems may generate outputs that have significant impacts on end-users or other parties, such as legal, financial, or health outcomes. They may generate code that violates laws, regulations, or ethical norms, or that causes harms or damages to end-users or their stakeholders. This may raise questions about the accountability and responsibility for the outputs and their consequences, and the allocation of liability and compensation in case of harms or damages.

\textbf{Attribution and ownership.} Generative AI models can create outputs that are similar or identical to existing human-created artefacts, such as code, images, text, etc. This raises questions about the attribution and ownership of the generated outputs, and the potential for plagiarism, infringement, or misuse of intellectual property \cite{sarkar2023exploring}. End-user programmers may not be aware of the original sources of the generated code, or the legal and ethical implications of using it for their own purposes.


In response to these challenges, researchers have proposed several complementary solutions \cite{hbr2020, harrer2023attention}:

\textbf{Ethical design.} Generative AI systems should be designed and developed with ethical principles and values in mind, such as fairness, transparency, explainability, privacy, security, and human dignity. Tools should incorporate mechanisms for data protection, output verification, bias mitigation, and user feedback.

\textbf{Critical use.} Generative AI systems should be used critically, and deployed with ethical standards and guidelines in mind, such as professional codes of conduct, industry best practices, and regulatory frameworks. End-user programmers need to be aware of the potential risks and challenges of using such tools.


\textbf{Governance.} Generative AI systems should be subject to ethical oversight and governance by various stakeholders, such as developers, providers, users, regulators, auditors, and ethicists. Tools should be monitored and evaluated by independent third parties for its performance, quality, safety, reliability, and accountability.

\section{Implications for End-User Programming Research}

We've explored how the generative shift results in a renewed research agenda around the role of formal systems in end-user programming, with a particular focus on control, agency, explanation and debugging, and awareness of the possible. This contrasts with two of the three main foci of prior generations of EUP research (described in Section~\ref{sec:eup-status-quo}): improving the ability of EUPs to learn formal systems, and to reduce the expertise requirements of formal systems. What might a research agenda for end-user programming look like in the wake of the generative shift?


\textbf{Learning to fish.} The importance of learning how to code for end-user programmers in a world where generative AI can provide answers is a complex issue that parallels the allegory of teaching a man to fish: \emph{``Give a man a fish, and you feed him for a day. Teach a man to fish, and you feed him for a lifetime''}. While the availability of generative AI systems that can provide direct answers may initially seem to diminish the need for coding skills, the underlying implications reveal a more nuanced perspective. Coding goes beyond the mere act of obtaining answers; it fosters self-sufficiency and cultivates a deeper understanding of the underlying processes and algorithms involved. By learning how to code, end-user programmers gain the ability to create their own solutions, adapt existing ones, and address unique challenges effectively. This empowerment leads to a greater sense of control and creativity, enabling them to solve problems beyond the scope of what AI systems can provide. It has been argued that coding skills facilitate critical thinking, problem-solving, and logical reasoning, which are valuable assets in various domains \cite{wing2006computational}. While generative AI can offer immediate solutions, relying solely on it can result in dependency and limit the potential for innovation. Therefore, while generative AI presents opportunities for efficiency and convenience, the significance of learning how to code persists, promoting independence, adaptability, and a deeper understanding of computational principles.

But is this really true of end-user programming? Is self-sufficiency really an important optimisation goal, if generative AI is consistently capable of solving end-user tasks? That is, do we still need to teach the man to fish if he can simply receive fish on demand? It is possible that the answer is ``no''; as much previous research has shown, in many end-user programming tasks the user may be content to use ``incorrect'' programs as well as manual methods in order to make progress towards a task \cite{pandita2018no, blackwell2002first}. Unlike in a CS education or professional programming setting, the \emph{task comes first}, and any educational or critical thinking outcomes are viewed as secondary. There are also objections to the universalising viewpoints of computational thinking, and evidence that learning to code does not in itself develop computational thinking skills \cite{denning2017remaining}.

A related issue has been explored by Potthast et al. \cite{potthast2021dilemma} in the context of information retrieval: ``the dilemma of the direct answer''. The question is to what extent search engine technology is responsible for synthesising information on the web to address the user's particular query. They identify various trade-offs both on the user side (e.g., the cognitive workload of analysing retrieved documents versus the accuracy of the answer) and system-side (e.g., the organisation of information for automated processing versus human reading). In our current context, we might think of this as ``the dilemma of direct programming''. An emergent research agenda for end-user programming might therefore be to ask what user-side and system-side tradeoffs emerge as a consequence of the generative shift.


\textbf{Repositioning BERTology.} Generative AI-based research must be carefully designed and evaluated in order to make meaningful contributions to the scientific literature. An emerging pattern in research is to simply try a generative model and see what it can do, a kind of research sometimes referred to as ``BERTology'' \cite{rogers2021primer}. This approach is passive: implicitly positioning EUP researchers outside the development process. Moreover, the results are unlikely to be robust to further iteration and development of generative AI (already the term ``BERTology'' shows signs of ageing, as its name refers to a family of models that is no longer considered the state of the art). Rather, experimenting with generative AI and developing prompting strategies is a \emph{craft practice} that should be seen as a necessary part of a research investigation, with its own forms of rigour, that feeds into broader research questions with more enduring value. Part of this rigour will involve identifying a clear research question or hypothesis that is independent of the ``abilities'' of models, the careful selection of a generative model that is well-suited to the problem at hand, and ensuring that the work is reproducible and transparent by detailing the models, heuristics, hyperparameters, and prompt development methodology.


\textbf{End-user software customisation.}
A closely related set of concerns to end-user programming is end-user software customisation \cite{mackay1991triggers}. This encompasses a wide range of activities such as changing the default settings in an application, writing macros, and editing software source code \cite{morch1997three}, all with the aim of tailoring software to user needs. End-user customisation is viewed as challenging, with many expertise and motivational barriers, and consequently quite rare; in fact the tendency of people to accept ``default'' options is a widely documented and deployed phenomenon in behavioural science in domains ranging across organ donation, retirement savings, browser and search engine choices, and wireless encryption \cite{kesan2006setting, jachimowicz_duncan_weber_johnson_2019}. At the same time, researchers have acknowledged the empowering value of end-user software customization, and proposed various potential solutions for increasing the customizability of software \cite{petricek_2022, basman2016software, klokmose2015webstrates}.

The generative shift may change the landscape of end-user software customization substantially. Not only in reducing the barriers to scripting and automation, but also in the spectrum of potential automations, and the user's relationship with and attitude to automation and scripting. Petricek theorises that customizable software consists of a set of ``substrates'' \cite{petricek_2022}, which are programming notations which trade off expertise requirements versus the scope of change they enable. For instance, in Excel, formulas have low expertise requirements and allow for a low scope of change to Excel functionality, whereas VBA has comparatively higher expertise requirements but enables a greater scope of change. A hypothetically ``ideal'' substrate can take on a smooth gradient, so that the same programming notation or environment can be used to make extremely small-scope changes with correspondingly low expertise requirements, as well as broad changes with correspondingly higher expertise requirements. Generative AI may provide such a substrate: by mediating between expressions of user intent in natural language, and a wide variety of underlying technical infrastructures. For instance, a natural language query in a spreadsheet might be satisfied by a combination of changed settings, formulas, and macros. Through the unified interface of naturalistic language, the user may be able to smoothly span a broad scope of changes. 

But the implications go further: rather than the current model of feature-rich software applications where a number of use cases have been ``captured'' in code by expert software developers \cite{nouwens2018application}, future applications for creating and manipulating information artefacts may leave a large portion of development and customisation for end users. In such a future, what is the nature of the application? Is it a small set of core features that identifies a nucleus of concerns or type of information artefact that is a useful abstraction for knowledge workers, upon which they build? Or does the application disappear as an organisational principle for knowledge work, transitioning into artefact or process-oriented paradigms?

The cognitive dimensions of notations framework \cite{green1996usability} offers a vocabulary for evaluating the design tradeoffs made in programming languages. However, they may also be applied to the design decisions in end-user software customisability. The generative shift enables much greater end-user flexibility in this regard, and raises some interesting implications for the cognitive dimensions framework. For instance, the dimension \emph{viscosity} refers to how difficult it is to make small changes to the program. Consider the spreadsheet, an archetypical end-user programming application. The spreadsheet interface may have its own sources of viscosity, but the spreadsheet also has multiple options for scripting and automation, such as the formula language, definition of custom functions, and macro-style scripting. Each of these may bring their own sources of viscosity. Now, if the spreadsheet interface can be significantly customised through scripting, the user may be in a position to identify and ameliorate sources of viscosity in their own workflow. This leads to the user experience of viscosity in the interface being \emph{multifaceted}, and evolving \emph{dynamically} over time, rather than relatively fixed properties of a notation. A related set of concerns has been explored in depth by Jakubovic et al. \cite{jakubovic2023technical}.



\textbf{Quantifying end-user programmers.} In 2005, Scaffidi et al. estimated the number of end-user programmers using US labour statistics and extrapolating from the number of spreadsheet users \cite{scaffidi2005estimating}. This methodology works for as long as EUP activity is strongly associated with particular applications (such as spreadsheets), but the generative shift will enable EUP activities across many more applications and platforms, facilitating the breakdown of application boundaries \cite{nouwens2018application} and mitigating the pains of transitioning between tools in the ``toolbelt'' style of computing \cite{sumner1997evolution, sarkar2023should}.


\textbf{Attention investment.} Blackwell's Attention Investment Model explains the behaviour of programmers in deciding whether to automate something (e.g., write a script) or pursue a manual strategy \cite{blackwell2002first}. Just like a cost-benefit analysis, Blackwell's model posits that a programmer is likely to pursue automation if the (perceived) \emph{payoff}, in terms of attention units saved by automation, minus the \emph{investment} costs of automation, is greater than the \emph{cost} of following a manual strategy, when accounting for the \emph{risk} that the automation may not work. What happens to this model when the cost of automation is reduced to zero, or near zero? It is possible that rather than a trade-off between the costs and benefits of attention investment into automation, other factors will come to dominate the automation decision, such as agency and trust. Or perhaps there will still a cost-benefit tradeoff in terms of attention units, but these units will be spent (and saved) primarily on new categories of EUP activities that do not currently exist.


\textbf{Learning barriers.} Ko et al. defined six learning barriers for end-user programming systems: design, selection, coordination, use, understanding, and information \cite{ko2004six}. After the generative shift, will EUPs face the same learning barriers? For instance, the abstraction matching problem identified by Sarkar et al. \cite{sarkar2022programmingai} does not fit neatly into any of these categories. Perhaps new learning barriers will appear, some will disappear, and some barriers will remain but their nature will change.

\textbf{Self-efficacy.} Wiedenbeck et al. found that computer self-efficacy increases substantially during introductory programming courses \cite{wiedenbeck2004factors}. They propose that in order to create self-efficacy, students should practice tracing program execution, program comprehension, and manual writing of code. However, as a result of the generative shift, EUPs may gain self-efficacy not from the direct authoring and comprehension of code, but from effectively applying tools to solve their domain problems. Comprehending the underlying logic and structure of the AI-generated code may require different instructional strategies compared to tracing the logic of manually written programs. Rather than focusing on code directly, EUPs might learn how to explore the outputs and performance characteristics of AI-generated code as a way of critically analysing its behaviour and identifying potential limitations. This approach may help EUPs develop a mental model that incorporates an understanding of the strengths and weaknesses of generative AI and how it aligns with their programming goals.

\textbf{Naming.} Liblit et al. draw attention to (identifier) naming as a central conceptualisation mechanism and cognitive challenge in programming \cite{liblit2006cognitive}. They find that programmers employ cognitive strategies when selecting and using names, such as following lexical and morphological conventions to convey role information and using metaphors to facilitate productive inferences. Natural language grammars influence name usage, and debates arise around the polysemy of names and their literal versus metaphorical meanings. The authors posit that the linguistic sophistication of a programming language may change or shift the cognitive burdens of programmers. The generative shift introduces an additional layer of linguistic influence in the prompt language used to control the tools as well as the language the tools may emit (by way of code explanation, example outputs, clarification requests, etc.). While generative AI could alleviate cognitive burdens in code development by generating optimised and readable code, as well as providing concise explanations of code logic, it is unclear whether such code and explanations can replicate the linguistic conventions, metaphors, and domain knowledge necessary for managing complexity, and it is uncertain to what extent such considerations remain central to the activity of programming as less code is written and read directly.

\textbf{Live programming.} Tanimoto presented a highly influential framework for evaluating the degree of liveness in programming environments, based on how quickly and continuously they provide feedback to programmers about the execution of their code \cite{tanimoto2013perspective}. After the generative shift, live systems might need to account for different forms and sources of feedback, besides execution feedback. For example, generative AI could provide feedback to programmers about the quality, readability, or maintainability of their code, or suggest alternative or improved ways of writing their code \cite{jain2020integrating}. Moreover, generative AI could enable programmers to use natural language, gestures, or voice commands to express their intentions or goals, and the programming environment could generate or modify code accordingly. This could change the notion of liveness from being based on code editing and execution, to being based on goal specification and realization. Alternatively, generative AI could enable programmers to delegate some or all of their programming tasks to intelligent agents, and monitor or intervene in their work as needed. This could change the notion of liveness from being based on direct manipulation and control, to being based on supervision and guidance \cite{tanimoto2020multiagent}.







\section{Conclusion}

This essay has considered how generative AI might change the landscape of end-user programming and end-user programming research. The research agenda for end-user programming has so far focused on helping people learn and use formal systems (e.g., a programming language). However, generative AI stands to facilitate an intensification and extensification of end-user programming activities across many more applications and tasks, and users may interact with generative AI primarily through informal systems of natural language. This is the \emph{generative shift hypothesis}.

Under the generative shift hypothesis, we have discussed how several strengths of formal systems, including control, agency, explanation, debugging, trust, may become less relevant or change in nature. We propose that EUP research may need to shift its traditional focus on formal systems to new concerns of the practical use of generative AI.

The key takeaways are:

\begin{itemize}
    \item The generative shift moves the focus of end-user programming research from improving the learnability and expertise requirements of formal systems, to new issues of control, agency, explanation, debugging, and the awareness of the possible.
    \item The generative shift affects many core theories and concepts of end-user programming, such as end-user software customisation, the attention investment model, learning barriers, self-efficacy, and live programming. These may need to be revisited and revised as generative AI enables EUP to proceed in new ways and scales.
\end{itemize}

Petricek encourages us to question the fundamental assumptions of programming languages research and theory \cite{petricekprogramming}. The generative shift is an opportunity to propel EUP research forward, renewing and revitalising the importance of human-centric approaches to instructing computers, rather than unprogrammable AI tools that ultimately take freedom away from the user. As Blackwell calls for in \emph{Moral Codes} \cite{Blackwell2022Chapter}: \emph{``If computer users have access to appropriate notations - Moral Codes - they can use simple automation to make their lives less mechanical, rather than more. If computer interfaces are designed as notational spaces, they offer freedom and negotiation, even forms of social organisation, complex assemblies of intelligent decision making and deliberation, respecting the humans creating them, rather than pretending humans were not involved.''}

\bibliographystyle{ACM-Reference-Format}
\bibliography{references}

\appendix

\end{document}